# Room temperature Epitaxial Stabilization of a Tetragonal Phase in $A$RuO$_3$ ($A$=Ca, Sr) Thin Films

**Arturas Vailionis**
*Geballe Laboratory for Advanced Materials, Stanford University, Stanford, California, 94305 and X-ray Laboratory for Advanced Materials, Stanford Linear Accelerator Center, Menlo Park, California 94025*

**Wolter Siemons and Gertjan Koster**
*Geballe Laboratory for Advanced Materials, Stanford University, Stanford, California, 94305 and Faculty of Science and Technology and MESA+ Institute for Nanotechnology, University of Twente, 7500 AE, Enschede, The Netherlands.*

*We demonstrate that SrRuO$_3$ and CaRuO$_3$ thin films undergo a room temperature structural phase transition driven by the substrate imposed epitaxial biaxial strain. As tensile strain increases, $A$RuO$_3$ ($A$=Ca, Sr) films transform from the orthorhombic phase which is usually observed in bulk SrRuO$_3$ and CaRuO$_3$ at room temperature, into a tetragonal phase which in bulk samples is only stable at higher temperatures. More importantly, we show that the observed phenomenon strongly affects the electronic and magnetic properties of $A$RuO$_3$ thin films that are grown on different single crystal substrates which in turn offers the possibility to tune these properties.*

The perovskite SrRuO$_3$ (SRO) and CaRuO$_3$ (CRO) materials are attracting considerable interest due to their fascinating electric and magnetic properties that can be utilized in heterostructures and oxide-based novel devices. These materials are mainly used in the form of thin films epitaxially grown on single crystal substrates.[1-3] Due to their close crystal lattice match with a range of functional oxide materials, SRO and CRO thin films are usually grown coherently on such substrates as SrTiO$_3$, DyScO$_3$, GdScO$_3$, LaAlO$_3$, and NdGaO$_3$ and serve as a technologically important bottom layer with atomically smooth surfaces and interfaces.[4,5] The lattice mismatch, which is usually present during heteroepitaxial growth, introduces strain in the layer and, in order for the layer to grow coherently, changes the lattice constants such that the in-plane lattice parameters of the film match those of the underlying substrate. The $A$RuO$_3$ epitaxial layers are usually grown with out-of-plane [110] direction with [001] and [1-10] directions aligned in-plane. A schematic depiction of heteroepitaxial growth of SrRuO$_3$(110) layer on SrTiO$_3$(001) substrate is demonstrated in Figure 1. As can be seen from the Figure, SrRuO$_3$ in-plane [001] and [1-10] directions are aligned along the [100] and [010] directions of the cubic SrTiO$_3$ substrate. In such a layer configuration the lattice mismatch between the layer and the substrate alters not only the $a$, $b$ and $c$ lattice constants but also the angle $\gamma$ between [100] and [010] axes of the film's unit cell.[6,7] This additional degree of freedom in misfit accommodation results in a distorted orthorhombic unit cell in epitaxial SrRuO$_3$ layers grown on SrTiO$_3$ substrates and most certainly changes Ru-O-Ru bond angles and/or Ru-O bond lengths, which are known to affect electric and magnetic properties of the $A$RuO$_3$ material.[8,9]

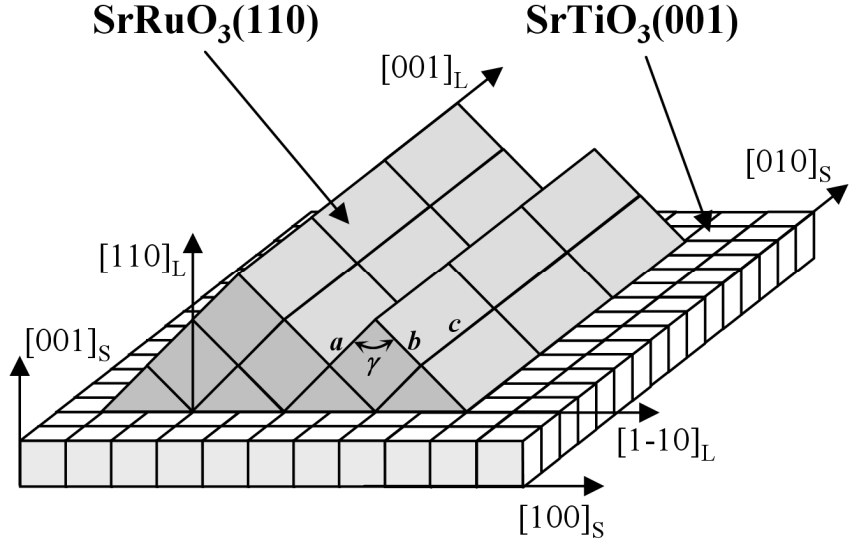

**SrRuO₃(110)**  **SrTiO₃(001)**

**Fig. 1**. Schematic representation of the heteroepitaxial growth of SrRuO$_3$(110) thin film on cubic SrTiO$_3$(001) substrate. The layer's [001] and [1-10] directions are aligned along [100] and [010] directions of the substrate.

From a technological point of view, it is therefore essential to learn how the unit cell parameters of $A$RuO$_3$ will vary with the change of applied biaxial stress induced by the different substrates. In this Letter we present a structural study of SrRuO$_3$ and CaRuO$_3$ thin films epitaxially grown on a number of single crystal substrates: SrTiO$_3$, DyScO$_3$, and NdGaO$_3$. The results show that the unit cell parameters of the $A$RuO$_3$ epitaxial layers are strongly affected by the strain sign and magnitude. Most importantly, we discovered that the tensile stress causes the $A$RuO$_3$(110) films to stabilize in a tetragonal instead of an orthorhombic structure at room temperature. From the crystallographic symmetry point of view, more symmetrical tetragonal unit cell is expected to significantly alter some of the Ru-O-Ru bond angles. We also demonstrate the effect of structural phase transition on the transport properties, which confirm that the electronic properties are deteriorating when the unit cell becomes more symmetric. Finally, we believe that our findings are more generally applicable to a large class of perovskite materials, such as the manganites, and offer the possibility to study the effect of unit cell symmetry on properties systematically.

SrRuO$_3$ and CaRuO$_3$ thin films were grown by Pulsed Laser Deposition (PLD). The samples were grown in the vacuum chamber with a background pressure of $10^{-8}$ Torr. All films were grown on TiO$_2$ terminated SrTiO$_3$ substrates.[10] Typical thicknesses of the films range from 20 to 30 nm. A 248 nm wavelength KrF excimer laser was employed with typical pulse lengths of 20-30 ns. The energy density on the target is kept at approximately 2.1 J/cm$^2$. Films were deposited with a laser repetition rate of 4 Hertz, with the substrate temperature at 700 °C in a mix of 50% Ar and 50% O$_2$ atmosphere with a total pressure of 320 mTorr. Note that the deposition conditions were kept constant for experiments on all substrate materials.

X-ray diffraction (XRD) measurements were performed using a PANalytical X'Pert materials research diffractometer in high- and medium-resolution modes at the Stanford Nanocharacterization Laboratory, Stanford University.

The XRD results demonstrate that epitaxially grown $A$RuO$_3$ films exhibit (110) out-of-plane orientation with (100) and (1-10) in-plane orientations. Reciprocal lattice maps from symmetrical and asymmetrical Bragg reflections allowed us to determine $a$, $b$, and $c$ lattice constants and $\alpha$, $\beta$, and $\gamma$ angles and therefore estimate the unit cell size and shape of the $A$RuO$_3$ layers. As an example, Figure 2 shows reciprocal lattice maps of the SrRuO$_3$ (260), (444), (620) and (44-4) reflections as well as the SrTiO$_3$(204) and DyScO$_3$ (260), (444), (620) and (44-4) reflections. As can be seen from Fig. 2(a), the SRO layer grown on STO substrate exhibits an orthorhombic unit cell symbolized by the difference in SrRuO$_3$ (260) and (620) atomic plane spacings which represent a dissimilarity between the $a$ and $b$ lattice parameters. In contrast, the tetragonal structure with $a = b$ shown in Fig. 1(b) for SRO film grown on DSO exhibit identical positions of (260) and (620) Bragg reflections.

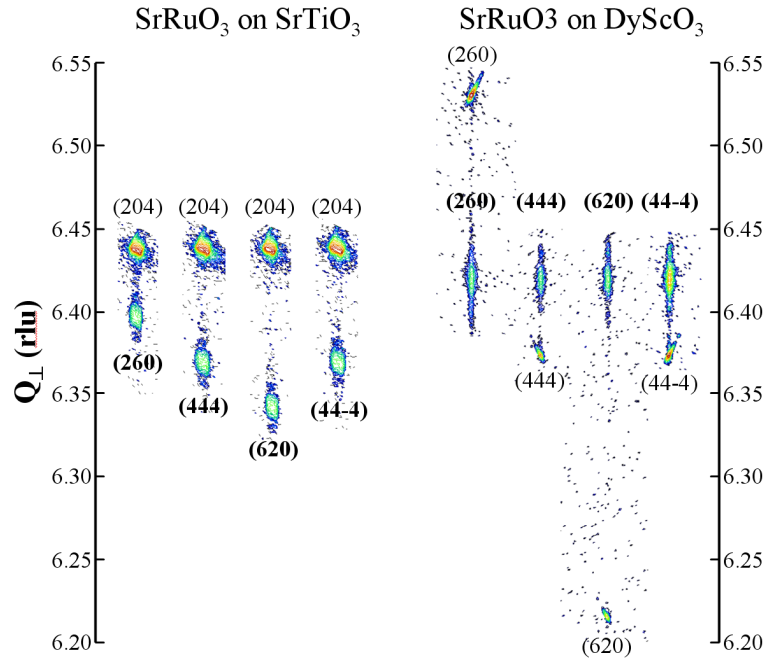

**Fig. 2**. Reciprocal lattice maps of SrRuO$_3$ layers grown on (a) SrTiO$_3$(001) and (b) DyScO$_3$(110) single crystal substrates. As can be seen, [1-10] and [001] in-plane lattice parameters for both layers are commensurate with the substrate's in-plane lattice parameters. SrRuO$_3$ on SrTiO$_3$ clearly shows an orthorhombic unit cell, while SrRuO$_3$ layer on DyScO$_3$ is tetragonal. Here we used $Q_\perp = 2\pi \sin\theta / \lambda$, where $\theta$ is the Bragg angle and $\lambda = 1.540598$ Å. Indexes of the layer's reflections are shown in bold.

The main structural results of the epitaxially grown $A$RuO$_3$ films are listed in Table I. As can be seen from the Table, all $A$RO layers under compressive stress demonstrate a $\gamma$ angle smaller than 90 degrees, while layers under tensile stress exhibit $\gamma$ angles larger than 90°. The $\gamma$ angle variation is consistent with the sign of the strain indicating that $A$RO unit cell, in addition to the variations of $a$, $b$, and $c$ lattice parameters, utilizes this additional degree of freedom to accommodate the mismatch between the substrate and the layer. As the negative mismatch becomes relatively large (SRO on DSO and CRO on STO) the $A$RO layer stabilizes in a tetragonal instead of an orthorhombic structure. The transition, which in bulk materials takes place at higher temperatures, in thin films is induced by epitaxial strain. It is important to note that orthorhombic-to-tetragonal (O-T) transition for SRO and CRO occurs at different mismatch values. As can be seen from Table I, CRO becomes tetragonal at a rather high mismatch values of about -1.78%, while for SRO O-T transition occurs at much lower mismatch of about -0.538%. The large dissimilarity can be explained by the different orthorhombicity factors (ratio of $a$ and $b$ lattice constants) of both materials. For bulk SRO orthorhombicity factor is 1.0066 much smaller than that of CRO which is 1.0318. The larger $a/b$ ratio in CRO material requires larger stress in order to switch its unit cell from orthorhombic to tetragonal.

**Table I**
Main structural parameters of SrRuO$_3$ and CaRuO$_3$ thin films grown on SrTiO$_3$(001), DyScO$_3$(110) and NdGaO$_3$(110) single crystal substrates. $\alpha$ and $\beta$ angles are equal to 90°.

| SrRuO$_3$ | | | | | | | |
|---|---|---|---|---|---|---|---|
| | Lattice parameters (Å and deg.) | | | | Strain (%) | | $a/b$ |
| Substrate | $a$ | $b$ | $c$ | $\gamma$ | [001] | [-110] | |
| SrTiO$_3$(001) | 5.529 | 5.577 | 7.810 | 89.41 | -0.441 | -0.439 | 1.0087 |
| DyScO$_3$(110) | 5.560 | 5.561 | 7.903 | 90.49 | 0.744 | 0.640 | 1.0002 |
| Bulk SrRuO$_3$ | 5.530 | 5.567 | 7.845 | 90.00 | - | - | 1.0066 |
| CaRuO$_3$ | | | | | | | |
| | Lattice parameters (Å and deg.) | | | | Strain (%) | | $a/b$ |
| | $a$ | $b$ | $c$ | $\gamma$ | [001] | [-110] | |
| NdGaO$_3$(110) | 5.359 | 5.535 | 7.706 | 90.28 | 0.745 | 0.392 | 1.0328 |
| SrTiO$_3$(001) | 5.461 | 5.463 | 7.760 | 90.42 | 1.451 | 0.778 | 1.0004 |
| Bulk CaRuO$_3$ | 5.354 | 5.524 | 7.649 | 90.00 | - | - | 1.0318 |

The strain imposed by the substrate changes the $a$, $b$, and $c$ lattice constants and the $\gamma$ angle and, as a result, modifies the Ru-O and Sr/Ca-O bond lengths and/or Ru-O-Ru bond angles. Due to this bond variation some physical properties of $A$RuO$_3$ are expected to be different for films grown on different substrates. The most noticeable change should be observed between orthorhombic and tetragonal $A$RuO$_3$ samples. It is known that for bulk SrRuO$_3$ the orthorhombic-to-tetragonal transition changes not only Ru-O bond lengths but also Ru-O-Ru bond angles.[11] The orthorhombic phase shown in Fig. 3(a) can be obtained by rotation of RuO$_6$ octahedra counterclockwise about the [010]$_{cubic}$ and [001]$_{cubic}$ directions and clockwise rotation about the [100]$_{cubic}$ direction of an $A$RuO$_3$ cubic perovskite.[11] Such rotations result in Ru-O-Ru bond angles being less than 180 degrees. Therefore, in the orthorhombic $A$RuO$_3$, the smaller bond angles between Ru and O are present in both $ab$ plane and along $c$-direction (apical oxygens). On the other hand, in the tetragonal unit cell shown in Fig. 3(b), RuO$_6$ octahedra are rotated only about the $[001]_{cubic}$ $A$RuO$_3$ direction. In this structural state bond angles between Ru and apical oxygens are equal to 180 degrees, while Ru-O-Ru bond angles in the $ab$ plane are less than 180 degrees. For strained SRO and CRO orthorhombic and tetragonal thin films, the actual Ru-O-Ru bond angles and Ru-O bond lengths are not known and currently are under investigation.

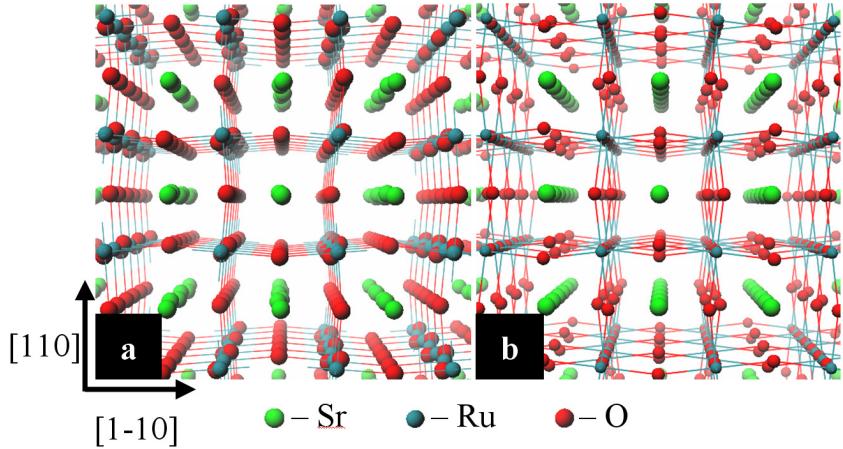

**Fig. 3.** Schematic representation of (a) orthorhombic (a ≠ b ≠ c) and (b) tetragonal (a = b ≠ c) structures of SrRuO$_3$. In the orthorhombic SRO structure, RuO$_6$ octahedra are rotated counterclockwise about [010] and [001] directions and clockwise about [100] direction. In the tetragonal structure oxygen atoms are rotated within (1-10) plane which results in rotations of RuO$_6$ octahedra around [001] direction.

A variation of the Ru-O-Ru bond angles most likely will affect the one electron bandwidth and will influence electron correlation and transport properties. As was shown recently, thin SrRuO$_3$ films with larger unit cell volume exhibit higher room temperature resistivity values due to ruthenium deficiency.[12,13] A similar effect could be the cause of the difference in transport properties measured for the samples in this study, but instead of the amount of ruthenium vacancies, strain lies at the heart of the differences observed here.

Typical resistivity measurements as a function of temperature for SRO and CRO films grown on different substrates are shown in Fig. 4. Thin layers that undergo large tensile in-plane strain (SRO on DSO and CRO on STO) have higher room temperature resistivities and their residual resistivity ratios (defined as the resistivity at 4 K divided by the one at 300 K) are lower than their less strained or compressively strained counterparts. The results would suggest that electron correlation in SRO and CRO is enhanced when a tetragonal unit cell is formed due to tensile strain. The increased correlation could as well be caused by a decrease of the one electron bandwidth or, equivalently, a reduction of the Ru-O-Ru bond angle.

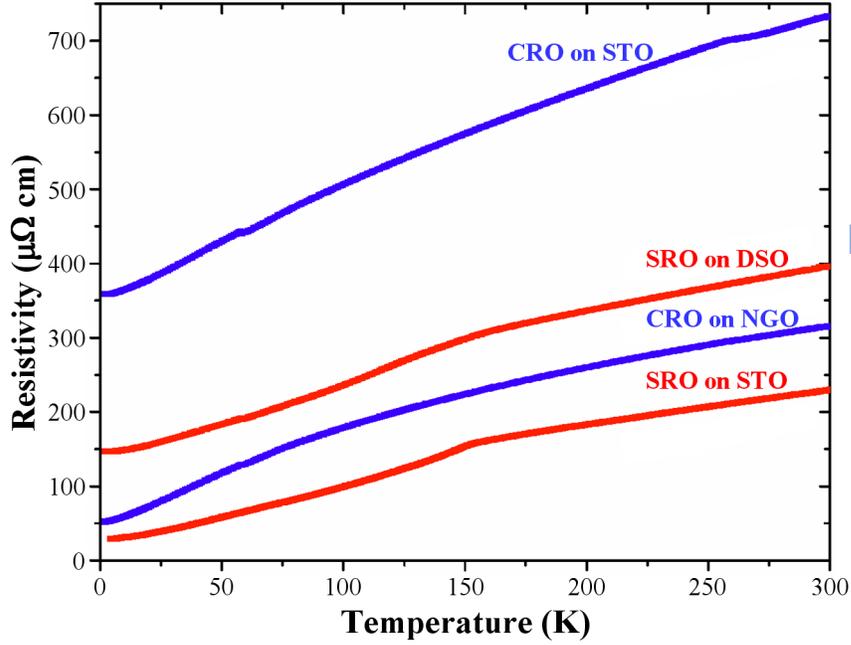

**Fig. 4.** Resistivity as a function of temperature of SrRuO$_3$ and CaRuO$_3$ thin films grown on different substrates. SRO on DSO and CRO on STO layers affected by a higher tensile strain are tetragonal and exhibit larger resistivity values, while orthorhombic SRO on STO and CRO on NGO layers show smaller resisitivity values.

In summary, we have demonstrated that the orthorhombic to tetragonal phase transition in SrRuO$_3$(110) and CaRuO$_3$(110) thin films can be induced by the biaxial stress. At lower mismatch values $A$RuO$_3$ films maintain orthorhombic unit cell with the angle between [100] and [010] directions, $\gamma$, being less than 90°. At higher mismatch values the $A$RuO$_3$ unit cell structure is tetragonal at room temperature. The higher symmetry tetragonal unit cell affects Ru-O-Ru bond angles and as a result deteriorates electronic properties of $A$RuO$_3$ ($A$ = Sr, Ca) thin films which are used in variety of technologically important applications.

This work was carried out under DoE BES support. Portions of this research were carried out at the Stanford Synchrotron Radiation Laboratory, a national user facility operated by Stanford University on behalf of the U.S. Department of Energy, Office of Basic Energy Sciences. The authors wish to acknowledge helpful discussion with J. Reiner, G. Rijnders, D.H.A. Blank, M.R. Beasley, T.H. Geballe and R.H. Hammond. W.S. thanks the Nanotechnology network in the Netherlands, NanoNed. One of us (G.K.) thanks the Netherlands Organization for Scientific research (NWO, VENI).